\documentclass{aa}    

\def\simless{\mathbin{\lower 3pt\hbox
     {$\rlap{\raise 5pt\hbox{$\char'074$}}\mathchar"7218$}}} 
\def\simgreat{\mathbin{\lower 3pt\hbox
     {$\rlap{\raise 5pt\hbox{$\char'076$}}\mathchar"7218$}}} 

\newcommand{\Msun} {M$_\odot$}

\usepackage {psfig, epsfig}

\begin{document}

\headnote{Letter to the Editor} 

\title{Discovery of 10~$\mu$m silicate emission in quasars.}
\subtitle{Evidence of the AGN unification scheme.}

\author {R.~Siebenmorgen\inst{1}
  \and M.~Haas\inst{2}
  \and E.~Kr\"ugel\inst{3}
  \and B.~Schulz\inst{4}
}

\institute{European Southern Observatory, Karl-Schwarzschildstr. 2, 
        D-85748 Garching b. M\"unchen
\and
        Astronomisches Institut, Ruhr-Universit\"at, 
        Universit\"atsstr. 150, D-44780   Bochum
\and 
        Max-Planck-Institut f\"ur Radioastronomie, Auf dem H\"ugel 69,
        Postfach 2024, D-53010 Bonn
\and 
        IPAC, California Institute of Technology (Caltech), Pasadena, 
        CA 91125, USA}

\offprints{rsiebenm@eso.org} \date{15.03.2005/10.04.2005}

\abstract{According to the unified scheme, AGN are surrounded by a dust-torus,
and the observed diversity of AGN properties results from the
different orientations relative to our line of sight.  The strong
resonance of silicate dust at 10 $\mu$m is therefore, as expected,
seen in absorption towards many type-2 AGN.  In type-1 AGN, it should
be seen in emission because the hot inner surface of the dust torus
becomes visible. However, this has not been observed so far, thus
challenging the unification scheme or leading to exotic modifications
of the dust-torus model.  Here we report the discovery of the
10~$\mu$m silicate feature in emission in two luminous quasars with
the Infrared Spectrograph of the Spitzer Space Telescope.

	}

\maketitle

\section{Introduction}

Observational evidence is mounting that most, if not all, quasars are
surrounded by large quantities of dust (Sanders et al. 1989, Haas et
al. 2003, Haas et al. 2004, Siebenmorgen et al. 2004).  According to
the AGN unification model, a quasar, i.e. a type-1 AGN, is seen
roughly pole-on allowing for a direct view of the nucleus, while a
type-2 AGN is seen edge-on with most of the central region being
hidden by the obscuring dust (Antonucci 1993).  The most convincing
support of the unification comes from spectro-polarimetric
observations in those edge-on cases, where scattering particles
located above and below the dust torus allow for viewing the
AGN-typical broad spectral lines of the central region (Heisler et
al. 1997).  A further argument in favour of unification comes from the
analysis of the isotropic far-infrared emission.  The fraction of the
nuclear luminosity that is absorbed by dust must be re-radiated in the
infrared, and sensitive observations with the ISO satellite found
similar far-infrared dust luminosities for radio loud type-1 and
type-2 AGN, after normalisation by their likewise isotropic 178 MHz
radio power. Hence, apart from possible starburst contributions and
other small differences, which may be caused by the various
evolutionary stages of these complex sources, the far-infrared
observations corroborate the unified scheme (Meisenheimer et al. 2001,
Haas et al. 2004).

Besides solid carbon, silicate minerals are a major component of dust
not only in the Milky Way, but probably also around AGN.  For a
centrally heated optically thick torus viewed edge-on, the 10~$\mu$m
silicate feature should be in absorption and, indeed, most type-2 AGN
display such an extinction signature.  On the other hand, when viewed
face-on, the hot illuminated surface of the inner torus wall should
display the silicates in emission.  This was already predicted in the
first studies of the dust radiative transfer in AGN tori (Pier \&
Krolik 1993).  For young stellar objects with surrounding disks, which
may be considered in many ways as scaled down versions of quasars, the
10~$\mu$m silicate emission is indeed also observed (Forrest et
al. 2004, van Boekel et al. 2004).  But contrary to the expectations,
the 10~$\mu$m spectra of type-1 AGN so far showed no clear indication
of the presence of silicates whatsoever, neither in emission nor in
absorption.  This poses a challenge to the AGN unification scheme.

\begin{figure*}
\hbox{\hspace{0cm}
\psfig{file=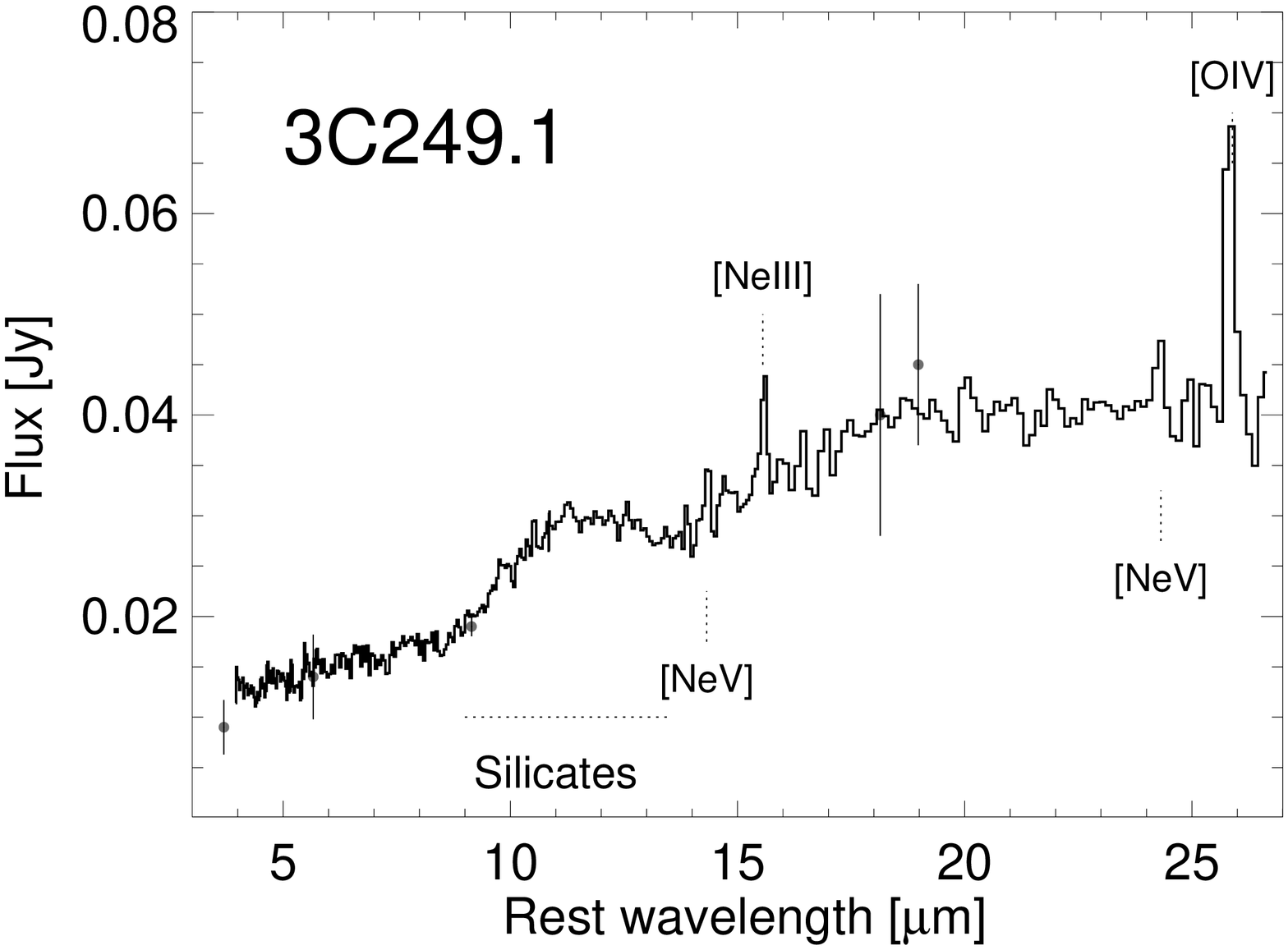,angle=90,width=9.cm,height=8cm}
\hspace{0.cm}
\psfig{file=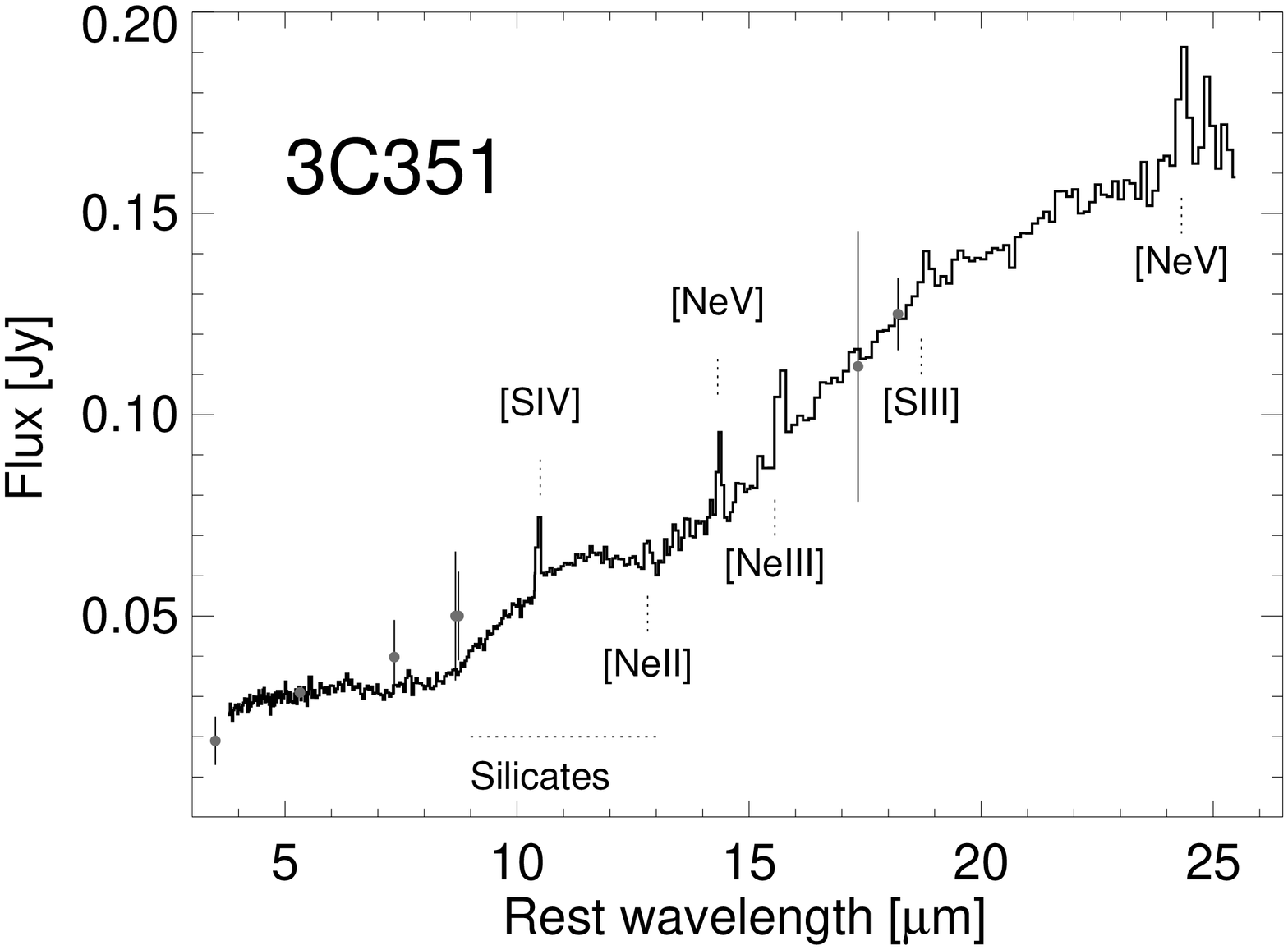,angle=90,width=9.cm,height=8cm}}
\caption{Spectra of the quasars 3C249.1 and 3C351, obtained with the 
Infrared Spectrograph (IRS) of the Spitzer Space Telescope. The
spectrum of 3C351 is scaled by a factor 1.1 in flux, to match the
broad band photometry from the literature (filled circles with
1$\sigma$-error bars). The AGN-typical high-excitation emission lines
like [\,Ne V\,] $\lambda$=14.3$\mu$m,$\lambda$=24.3$\mu$m are marked
with vertical dotted lines. The broad bump in the wavelength range
9 -- 13 $\mu$m, indicated by horizontal dotted lines, is the detected
silicate emission. \label {1.fig}}
\end{figure*}

For more than a decade, several research groups have therefore sought
for ways to reconcile the straightforward model predictions with the
observational fact of non-detection.  This led to severe modifications
of the dust torus models.  As solutions, it was proposed that the
silicon abundance in the AGN dust is very low, or that the grains are
much bigger than what is usually adopted (Laor \& Draine 1993).  Then
AGN would radiate like featureless blackbodies in the 10~$\mu$m
region, but for what reason should dust grains coagulate to larger
particles in the hostile environment of a luminous quasar?  In another
attempt, the geometry of the dust distribution was fine-tuned,
resulting in sophisticated shapes; tapered disk configurations were
presented in order to allow for a reduction of the observable silicate
emission up to large, close to face-on, viewing angles (Efstathiou \&
Rowan-Robinson 1995).  Rowan-Robinson (1995) suggested that an
ensemble of geometrically small but optically thick clouds would
result in only a weak feature, due to the effective cancellation of
absorption and emission processes in the 10~$\mu$m line. Similarly,
clumped dust density distributions were investigated (Nenkova et
al. 2002), however, the claimed reduction of the silicate 10~$\mu$m
emission has recently been questioned (Dullemond \& van Bemmel 2005).

\section{Observations}

Because the missing silicate emission presents a riddle of possibly
fundamental importance, we observed a number of steep radio spectrum
quasars and powerful radio galaxies using the infrared spectrograph
(IRS) of the Spitzer Space Telescope (Werner et al. 2004).  The sample
contains the two luminous quasars 3C249.1 (PG 1100+772) and 3C351 (PG
1704+608), at redshift z = 0.312 and z = 0.372, respectively.  They
have a blue luminosity M$_{\rm B}$ $\sim$ -26 mag, and an infrared
dust luminosity estimated from ISO observations to be 4 $\times$
10$^{\rm 45}$ erg/s assuming a $\Lambda$-cosmology with H$_0$ = 71 km
s$^{\rm -1}$Mpc$^{\rm -1}$, $\Omega$$_{\rm m}$ = 0.27 and
$\Omega$$_{\rm \lambda}$ = 0.73.  ROSAT and ASCA X-ray data imply that
the intrinsic absorption is negligible (Sambruna et al. 1999, Brandt
et al. 2000).  Hence, both sources are classified as powerful type-1
AGN with an almost unobscured line-of-sight towards the hot inner wall
of the putative dusty torus.  Therefore, these two quasars are well
suited to search for the 10~$\mu$m silicate emission feature.

The objects were observed between 5 and 35 $\mu$m in the two IRS low-resolution
(64 $<$ $\lambda$ / $\Delta$$\lambda$ $<$ 128) modules in staring mode.  Our
analysis starts with the two dimensional data frames from the Spitzer pipeline
(Higdon et al. 2004) using the latest calibration files.  At this point the
major instrumental effects are already removed.  We subtracted the sky
background, using pairs of frames, where the source appears at two different
positions along the spectrometer slit.  We interactively extracted the one
dimensional spectra.  Before averaging, parts with low reproducibility between
integrations were discarded.

\begin{figure*}
\hbox{\hspace{0cm}
\psfig{file=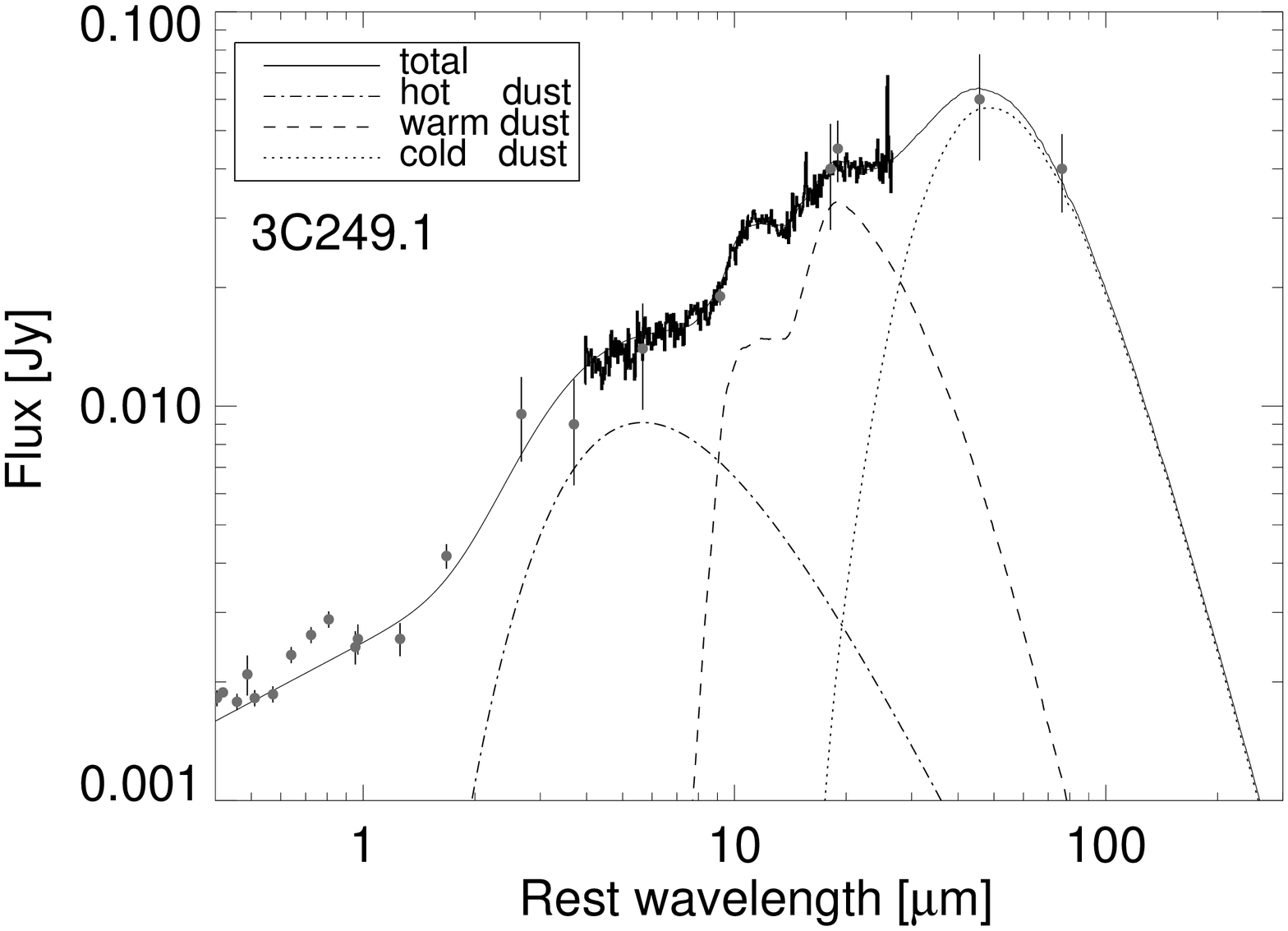,angle=90,width=9.cm,height=8.5cm}
\hspace{0.cm}
\psfig{file=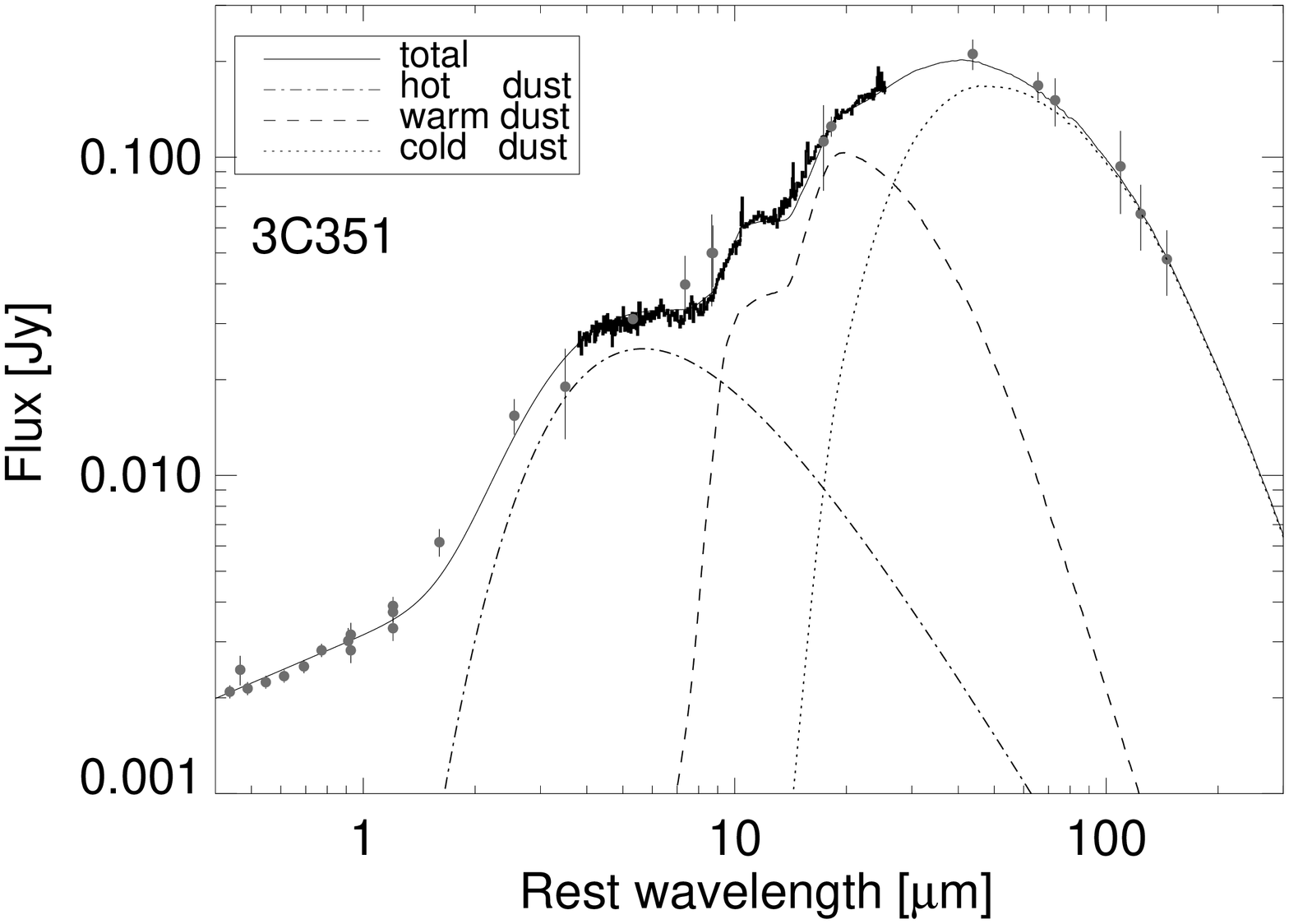,angle=90,width=9.cm,height=8.5cm}}
\caption{
  Spectral data (see Fig.\ref{1.fig}) and model fit for the quasars
  3C249.1 and 3C351. The models consist of a central heating source
  with an X-ray-to-infrared power-law spectrum, two dust components
  and a black body, their parameters being given in Table~\ref{1.tab}.
 }
\label {2.fig}
\end{figure*}
\begin{table*}
\caption{Model components of the infrared emission\label{1.tab}.
In both quasars, the infrared emission can be fitted by a
superposition of a blackbody and a warm and a cold dust component (see
Fig.~\ref{2.fig}). The temperatures, the surface area of the
blackbody, the mass of the dust components and their distance assuming
direct heating by the AGN are listed. For better agreement with the
data, we used for 3C351 two warm and two cold components. }
\centerline{
{
\begin{tabular}{l|cc |ccc|ccc}
\hline
  & & & & & & & & \\
\multicolumn{1}{c}{Name}      & 
\multicolumn{2}{ c}{Blackbody}  & 
\multicolumn{3}{ c}{Warm dust} & 
\multicolumn{3}{c}{Cold dust} \\
  & & & &   & & & & \\
\hline
  & & & &   & & & & \\
     & T   & Area & T & Mass & Dist. & T & Mass & Dist. \\
 & [K] &[pc$^2$] &[K] & [$10^3$ \Msun]  & [pc]& [K]& [$10^6$ \Msun] & [kpc] \\
  & & & &   & & & & \\
\hline
  & & & &   & & & & \\
3C249.1 & 900 & 180 & 180 & 3.3 & 30 & 60 & 1.4 & 1.0 \\
  & & & &   & & & & \\
\hline
  & & & &   & & & & \\
3C351   & 900 & 725 & 140 & 35 & 100 & 33 & 53 & 5.7 \\
        &     &     & 195 & 5.3 & 30 & 70 & 2.3 & 0.6 \\
  & & & &   & & & & \\
\hline
\hline
\end{tabular}
}
}
\end{table*}


\section{Results and Discussion}

Fig.~\ref{1.fig} depicts the IRS spectra shifted to the quasar rest
frames (4 -- 27 $\mu$m).  Not only do they show the AGN-typical
high-excitation lines like [\,Ne V\,] $\lambda$=14.3$\mu$m,
$\lambda$=24.3$\mu$m and [\,O IV\,] $\lambda$=25.9$\mu$m, but they also
exhibit a prominent broad bump in the wavelength range 9 -- 13
$\mu$m. By comparison with the IRS spectra obtained for other sources
of the sample (3CR radio galaxies), we rule out that the bump is an
artifact.  Furthermore, if it were produced by broadening of an atomic
line, the velocity dispersion would have to be unacceptably large,
about one fifth of the speed of light.  Infrared emission bands in
this wavelength range, such as features of polycyclic aromatic
hydrocarbons, do also not fit.  As the shape of the bump is
explainable by optically thin emission of silicate minerals, we
conclude that this is in fact their origin, and that we here detected
this long sought for emission feature in quasars. We find that in both
quasars the silicate emission contributes to about 20\% of the total
9 -- 13 $\mu$m luminosity.

 The absorption coefficient of interstellar silicate dust,
 $\kappa_{\nu}$, has a local maximum around 9.7$\mu$m producing the
 well-known broad band. The exact position depends on the choice of
 the optical constants. The emission feature detected here is centered
 around 11$\mu$m. Such a shift in the local maximum can occur as a
 result of varying grain mineralogy or an increase in grain size
 ($2\pi \cdot a/\lambda \sim 1$) as suggested for circumstellar matter
 for young stars (Bouwman et al. 2001, Forrest et. 2004). In our case,
 where the emission is probably optically thin and therefore
 proportional to $\kappa_{\nu} \times B_{\nu} (T_{d})$, the natural
 explanation for the local extrema is the folding of $\kappa_{\nu}$
 with the steeply rising Planck function $B_{\nu} (T_{d})$. The
 presence of carbon grains of about 180K whose emission is included in
 the observed flux enhances the effect and broadens the bump.

We can reproduce the Spitzer observations, as well as the photometric
data at other infrared wavelengths, by a model consisting of three
components: cold dust, warm dust and a hot blackbody. The primary
heating source has a power-law spectrum,
F$_{\nu}$\,$\propto$\,$\nu$$^{\rm -0.7}$, in the wavelength range
0.1~nm -- 15~$\mu$m.  We assume that the dust in the quasar is similar
to that of the Milky Way; it is a mixture of carbon and silicate
spheres of fixed radius of 0.1 $\mu$m with optical constants by Zubko
et al. (2004) and cross sections calculated from Mie-theory. The
emission of the warm and cold component is optically thin which is a
reasonable assumption for a face-on viewed quasar. The parameters of
the dust components, i.e. their temperature, mass and characteristic
distance to the AGN, are listed in Table~\ref{1.tab}, together with
the temperature and surface area of the blackbody. The emission of the
latter may also be due to dust, but hot and optically thick. As shown
in Fig.~\ref{2.fig}, this simple model fits the data and, in
particular, the silicate emission bump quite well suggesting that it
is not necessary to postulate a very low silicate abundance, exotic
grain sizes or sophisticated torus geometries. Of course, this
approach needs further refinement by a self-consistent axial-symmetric
radiative transfer model of the quasar emission. Such studies are in
progress and will be presented in a forthcoming paper.

Our detection of the silicate emission in two quasars poses the
question why it has not been seen so far in other type-1 AGN sources.
All previous spectroscopic observations which could suitably cover the
7 -- 15 $\mu$m rest frame wavelength range necessary to unambiguously
identify the broad emission bump, were of lower sensitivity and hence
restricted to nearby AGN (Sturm et al. 2002). These Seyfert-1 sources,
however, are intrinsically a factor between ten and hundred less
luminous than the two distant quasars investigated here.  Therefore,
our first guess is that the luminosity determines whether the silicate
emission band is prominent or not. Further investigations may provide
clues to a possible luminosity dependence of the silicate emission
strength.

\acknowledgements

This work is based on observations with the Spitzer Space Telescope
operated by the Jet Propulsion Laboratory, California Institute of
Technology, under contract with NASA. This research was supported by
the Nordrhein-Westf\"alische Akademie der Wissenschaften, funded by
the Federal State Nordrhein-Westfalen and the Federal Republic of
Germany. We thank the referee Michael Rowan-Robinson for his
constructive report.

\end{document}